\documentclass[acmsmall,screen]{acmart}
\AtBeginDocument{%
  }

\setcopyright{none}
\renewcommand\footnotetextcopyrightpermission[1]{} 
\renewcommand\footnotetextcopyrightpermission[1]{} 
\AtBeginDocument{%
  
  \makeatletter
  
  \makeatother
  \fancyhf{} 
  \fancyfoot[C]{\thepage} 
  \pagestyle{fancy} 
}

\usepackage{adjustbox}
\usepackage{subcaption,booktabs,siunitx}
\usepackage{multirow}
\usepackage{array}
\usepackage[table]{xcolor}
\begin{document}

\title{Complexity at Scale: A Quantitative Analysis of an Alibaba Microservice Deployment}

\author{Giles Winchester}
\email{G.Winchester@sussex.ac.uk}
\orcid{0000-0002-5928-228X}
\affiliation{%
  \institution{University of Sussex}
  \city{Brighton}
  \country{UK}
}

\author{George Parisis}
\email{G.Parisis@sussex.ac.uk}
\orcid{0000-0002-1298-7143}
\affiliation{%
  \institution{University of Sussex}
  \city{Brighton}
  \country{UK}
}

\author{Guoyao Xu}
\email{yao.xgy@alibaba-inc.com}
\orcid{0000-0002-1136-2678}
\affiliation{%
  \institution{Alibaba}
  \city{Hangzhou}
  \country{China}
}

\author{Luc Berthouze}
\email{L.Berthouze@sussex.ac.uk}
\orcid{0000-0003-3774-2369}
\affiliation{%
  \institution{University of Sussex}
  \city{Brighton}
  \country{UK}
}


\begin{abstract}
Microservice management and testbed research often rests on assumptions about deployments that have rarely been validated at production scale. While recent studies have begun to characterise production microservice deployments, they are often limited in breadth, do not compare findings across deployments, and lack consideration of the implications of findings for commonly held assumptions. We analyse a distributed tracing dataset from Alibaba's production microservice deployment to examine its scale, heterogeneity, and dynamicity. By comparing our findings to prior measurements of Meta's MSA we illustrate both convergent and divergent properties, clarifying which patterns may generalise. Our study reveals extreme architectural scale, long-tail distributions of workloads and dependencies, highly diverse functionality, substantial call graph variability, and pronounced time-varying behaviour which diverge from assumptions underlying research models and testbeds. We summarise how these observations challenge common assumptions in research on fault management, scaling, and testbed design, and outline recommendations for more realistic future approaches and evaluations.

\end{abstract}



\maketitle
\thispagestyle{empty}

\section{Introduction}

Microservice architectures (MSAs) have become the leading method for building large-scale distributed applications \cite{livens_what_2023}. These architectures are built using distinct microservices (MSs) that communicate with each other over well defined APIs to provide application logic. This architectural style facilitates small agile development teams, rapid development of new features, and fine-grained scaling. Yet, these characteristics can also lead to unique challenges associated with MSAs, such as increased operational complexity, MS co-ordination, and scale \cite{soldani_pains_2018}. This renders system maintenance, integrity, and uptime assurance extremely complicated. However, on the other hand, online application downtime can be associated with extremely high costs, with Fortune 1000 companies reporting losses of up to \$1 million per hour of disruption \cite{atlassian_calculating_nodate, noauthor_cost_nodate}.

To address these challenges extensive research has been carried out into novel approaches to detect faults \cite{nobre_anomaly_2023}, pinpoint root causes \cite{zhang_failure_nodate}, provide efficient auto-scaling through workload prediction \cite{luo_power_2022}, and optimise MSA design by minimising anti-patterns \cite{gamage_using_2021, parker_visualizing_2023}. However, these approaches are often motivated and subsequently designed based on theoretical assumptions of MSAs stemming from perceived common designed practices and the challenges associated with them. This issue is further exacerbated by the reliance on small-scale open-source testbeds \cite{seshagiri_sok_2022} due to the scarce access to industrial-scale datasets with ground truth. Whilst such testbeds are designed based on best practices, they do not match the complexity associated with real-world large-scale MSAs \cite{seshagiri_sok_2022}. 

Previous work has begun to characterise the attributes of large-scale MSAs, such as the properties of MS call graphs \cite{luo_characterizing_2021, wen_bytedance_2022}, runtime performance \cite{luo_characterizing_2021, wen_bytedance_2022}, and the time-varying evolution of dependencies at runtime \cite{winchester_temporal_2023}. However, these studies only consider characteristics from a narrow viewpoint. To address this limitation, a recent study was carried out on the characteristics of Meta's MSA \cite{huye_lifting_2023}, taking a broader view of MSA characteristics. However, this analysis still omitted several important characteristics of MSAs and the implications of their findings for
the assumptions leveraged by researchers are not discussed in depth. Furthermore, without a substantial comparison to other large-scale deployments, it remains unclear how representative their results are of MSAs as a whole. Despite the growing literature, there still lacks a broad, empirically grounded understanding of how operational complexity manifests at scale, and whether observations at one provider generalise to another.

To address this gap, we make the following three contributions: 

\begin{enumerate}
    \item \textbf{Comprehensive, multi-dimensional analysis:} building on prior work that identifies scale, heterogeneity, and dynamicity as the primary sources of operational complexity in large-scale MSAs \cite{soldani_pains_2018, seshagiri_sok_2022, zhou_fault_2021, gan_sleuth_2024, gan_sage_2021,lin_log_2016, yang_robust_2022, lin_microscope_2018}, we carry out a comprehensive, multi-dimensional analysis of Alibaba's deployment across these three axes, offering a broader view than prior Alibaba and Meta studies.
    \item \textbf{Cross-deployment comparison:} we provide the first cross-deployment comparison of two large-scale MSAs, revealing convergent and divergent properties.
    \item \textbf{Operational insights and research implications:} we highlight and discuss how real-world MSA operational properties challenge common assumptions in management approaches and testbed design, and we offer recommendations (summarised in Table \ref{tab:assumptions_findings}) to guide future work to better meet the challenges of real-world MSA deployments.
\end{enumerate}

All code and the processed data used to generate the figures and statistics in this paper will be publicly released upon acceptance of the paper.


\section{Related Work}

Quantitative studies of real-world MSAs are relatively few and narrow in scope, though several notable works exist. Luo et al. \cite{luo_characterizing_2021} analysed call graph characteristics from 12 hours of Alibaba traces, reporting depth and complexity statistics and clustering call graphs based on topology. Winchester et al. \cite{winchester_temporal_2023} examined the same data, focussing instead on the time-varying evolution of global dependency structure, revealing repeating system-wide states that could be linked to global system performance. While both studies offered valuable insights, their reliance on a short 12-hour window limits confidence in the completeness of their findings.

Luo et al. \cite{luo_power_2022} later analysed 14 days of Alibaba traces to study time-varying workload in relation to dependency dynamics, topological characteristics, and architectural roles. Despite the larger dataset, this study, similar to previous work, also concentrated on a limited set of characteristics, providing only a partial view of the broader properties of large-scale MSAs.

More recently, Huye et al. \cite{huye_lifting_2023} conducted a broader characterisation of Meta's MSA, analysing its dependency topology, architectural evolution, and depth and variability of traces for a select few front-end services. Although this work presented a more comprehensive view than previous studies, it does not explore important aspects of MSAs, such as workload, cohabitation, uncommon call graphs, and time-varying topology. Furthermore, this work does not explore the implications of their findings in relation to common assumptions held by MSA management approaches. Moreover, the lack of comparison with other deployments leaves it unclear whether the reported observations reflect common characteristics of large-scale MSAs or instead are specific to Meta's MSA.

These gaps motivate our comprehensive analysis of Alibaba's deployment and our cross-deployment comparison with Meta, to exhibit how operational characteristics of real-world MSAs challenge common assumptions in management and testbed design. 

\begin{table}[t]
\centering
\small
\caption{Example trace from the \textit{call data}.}
\label{tab:example-callgraph}
\begin{adjustbox}{max width=\linewidth}
\begin{tabular}{c c c c c c}
\toprule
\textbf{Timestamp} & \textbf{Trace ID} & \textbf{RPC ID} & \textbf{Service ID} & \textbf{Upstream MS} & \textbf{Downstream MS}\\
\midrule
115352 & T\_1156086 & 0.1   & S\_153587416 & MS\_58845 & MS\_71712 \\
115354 & T\_1156086 & 0.2   & S\_153587416 & MS\_58845 & MS\_72104 \\
115358 & T\_1156086 & 0.2.1 & S\_153587416 & MS\_72104 & MS\_90011 \\
\bottomrule
\end{tabular}
\end{adjustbox}
\end{table}

\begin{table}[t]
\centering
\small
\caption{Example rows from the \textit{resource data}.}
\label{tab:example-metric}
\begin{adjustbox}{max width=\linewidth}
\begin{tabular}{c c c c c c}
\toprule
\textbf{Timestamp} & \textbf{MS ID} & \textbf{Instance ID} & \textbf{Node ID} & \textbf{CPU Util.} & \textbf{Mem. Util.} \\
\midrule
180000 & MS\_21881 & MS\_21881\_POD\_0 & NODE\_11517 & 0.21996 & 0.83300 \\
180060 & MS\_21881 & MS\_21881\_POD\_0 & NODE\_11517 & 0.19847 & 0.81052 \\
180120 & MS\_21881 & MS\_21881\_POD\_1 & NODE\_11519 & 0.24216 & 0.85499 \\
\bottomrule
\end{tabular}
\end{adjustbox}
\end{table}

\section{Alibaba's Microservice Architecture and Dataset}
\label{sec:dataset}

The Alibaba MSA has been previously described in \cite{luo_power_2022}, for this paper we provide a brief overview. Alibaba have adopted Kubernetes \cite{noauthor_production-grade_nodate} to manage their cloud \cite{zhang_high-density_2020}. Stateful MSs, such as databases and in-memory stores, are deployed within a dedicated cluster which is not shared with stateless MSs. Alibaba employ the Application Real-Time Monitoring Service (ARMS) \cite{noauthor_application_nodate, cai_real-time_2019} to collect monitoring information from their deployment. The data collected by ARMS over 14 days of operation of deployed MSs in 2022 has been open-sourced \cite{noauthor_alibabaclusterdata_2025}. The dataset is split into several parts, each covering specific monitoring information; in this paper, we utilise the data associated with MS calls and resource usage, from hereafter referred to as the \textit{call data} and \textit{resource data}. Certain deployment-specific details that are not publicly available in the description of the dataset and findings from analysis were clarified via private communication with the dataset's maintainers.

Each row of the \textit{call data}, shown in Table \ref{tab:example-callgraph}, corresponds to a single call made between an upstream and downstream MS, acting as the consumer and provider MSs, respectively. Each row also contains a number of auxiliary fields. The \textbf{Trace ID} identifies the collection of calls made between MSs that are associated with a single front-end request. The \textbf{RPC ID} is used to reconstruct the call graph structure of a call collection identified by a given Trace ID by establishing the invocation order of calls starting from the entry MS, in a dot-separated format. For example, the two calls made by an entry-level MS \textit{A} to MSs \textit{B} and then \textit{C} will have the RPC IDs `0.1' and `0.2' respectively. If \textit{C} subsequently made a call to \textit{D} the RPC ID for that call would be `0.2.1'. Finally, the \textbf{Service ID} associates each call group within the data with the invocation of a specific front-end service functionality. These functionalities include searching products, ordering, and delivering\footnote{Service IDs are also associated with sporadic non-user functionality such as management and testing, and thus this interpretation of Service IDs is not completely straightforward, as seen in Section \ref{sec:scale_services}.}. However, these are anonymised in the dataset.

Each row of the \textit{resource data}, shown in Table \ref{tab:example-metric}, contains a single value of CPU and memory utilisation from stateless MS instances, sampled every 60 seconds across the 14 days. Each row contains a \textbf{MS Instance ID} and \textbf{MS ID} which correspond to the stateless MS names found in the \textit{call data}. Each row also contains a \textbf{Node ID} which denotes the specific host in which the MS instance is deployed at the time of the sample.

\begin{table*}[h]
  \caption{Counts of MSs and MSI by call‑pattern category: entry, leaf, middle, and total.}
  \label{tab:counts}
  \centering
  \footnotesize
  \setlength{\tabcolsep}{4pt}
  \begin{tabular}{lcccc}
    \toprule
               & Entry         & Leaf          & Middle        & Total     \\
    \midrule
    MS         & 8,591 \textbf{(13\%)}  & 25,201 \textbf{(39\%)} & 30,959 \textbf{(48\%)} & 67,760    \\
    MSI        & 137,547 \textbf{(7\%)} & 436,263 \textbf{(23\%)} & 1,292,281 \textbf{(69\%)} & 1,866,091 \\
    Ratio      & 16            & 17            & 41            & 28        \\
    \bottomrule
  \end{tabular}
\end{table*}

\section{Scale}
MSAs, whilst conceived for individual MS simplicity and modularity, can often evolve into highly complex, sprawling architectures as they organically grow to meet business demands, imparting significant operational challenges for MSA management \cite{soldani_pains_2018}. In this section, we present quantitative insights into how scale manifests at the architectural, service, and MS levels.

\subsection{Microservices and Instances}
\label{sec:scale_ms_msi}

\textbf{Scale is measured in tens of thousands of MSs and a million instances:} we identified 64,760 unique MSs that were deployed as 1,866,091 instances in 14 days (see Table \ref{tab:counts}); on average, each MS was deployed as 29 instances.

\textbf{Most MSs both make and receive calls:} we found that 8,591 MSs were responsible for handling entry calls\footnote{An entry call is the initial request made from an external client or system that enters the MS deployment.} and 25,201 of MSs never made calls across the 14 days, meaning that 30,959 of MSs, both received and made calls. Based on the type of RPC IDs of received calls we found that 61\% of leaf MSs were stateful. However, a noteworthy share of leaf MSs were stateless, illustrating that some MSs deliver functionality without any additional database calls.

\textbf{MS that both make and receive calls had more deployed instances:} across all MS instances we found that 728,544 (39\%) appeared downstream at least once, whereas 1,429,828 (76.6\%) appeared upstream. MSs that only received calls made up a lower than expected share of MS instances -- 34.3\% of MSs but only 23.4\% of instances. This finding may be attributed to reduced scaling due to higher horizontal scaling overheads from the need for data replication, consistency, and distributed transaction management. Overall, MSs that both received and made calls comprised the majority of MS instances, 69.3\%, indicating that these MSs experience higher or more varied workloads, or their functionality is associated with higher computational costs. 

\textbf{Alibaba's deployment is larger in number of unique MSs but Meta's is larger in number of replications:} whilst Meta had fewer unique MSs (18,500 vs 64,800) these MSs were deployed as considerably more instances than at Alibaba (12 million vs 1.8 million). This suggests that Meta's architecture may consolidate broader functionality into fewer MSs that are scaled out more aggressively, whereas Alibaba's appears to decompose functionality into a larger number of fine-grained MSs. While instance replica count is influenced by many factors such as workload intensity, geo-distribution, and resilience targets, these findings hint at different architectural choices. This divergence likely not only reflects the contrasting service domains of Meta (social media centric) and Alibaba (e-commerce, financial, and digital media) but also differing scaling and operational constraints.

\begin{figure}[h]
  \centering
  \includegraphics[width=0.55\columnwidth]{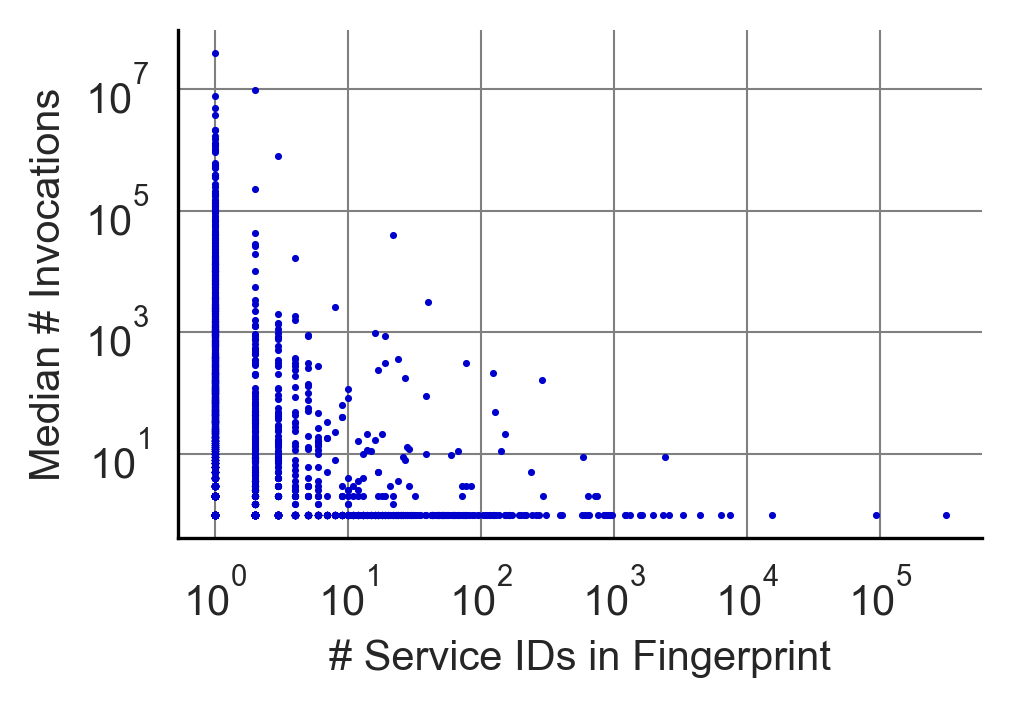}
  \caption{Service ID counts vs. the median number of invocations per \textit{call fingerprint}.}
  \label{fig:scale_fingerprints_invocations}
\end{figure}

\subsection{Services}
\label{sec:scale_services}

\textbf{The number of Service IDs is significantly inflated by sporadic non-user functionality:} we identified 166,093,303 unique Service IDs, dwarfing the number of MSs. However, Service IDs can also be associated with non-user functionality. When investigating the number of invocations of each service across the 14 days we found that only 7,352,480 (4.4\%) had more than a single invocation, and only 536,958 (0.3\%) had more than 100 total invocations. However, these values are inflated by non-user facing functionality (see \ref{sec:dataset}). This is reinforced by the fact that Alibaba's online services are generally used by a very large number of users, thus a large number of services having very few invocations unrealistic.

\textbf{Many Service IDs share an identical set of MS call repertoires:} for each Service ID, we grouped all associated MS calls into a deduplicated list. This list represents all of the MSs that the front-end functionality relies on, which we refer to as a Service ID's \textit{call fingerprint}. We explored to what extent different Service IDs had overlapping \textit{call fingerprints}, and thus overlapping functionality. We converted \textit{call fingerprints} into MinHashes and used locality-sensitive hashing (LSH) to match fingerprints. Overall, we found that 159 million (99.39\%) Service ID \textit{call fingerprints} could be matched to at least one other \textit{call fingerprint}, and 10\% of all Service IDs shared the same \textit{call fingerprint}.

\textbf{Sporadic Service IDs have \textit{call fingerprints} that are shared more:} for a random sample of 10,000 \textit{call fingerprints} we grouped Service IDs associated with each \textit{call fingerprint} and calculated the median invocations. We found that the more a \textit{call fingerprint} was shared among Service IDs, the fewer invocations each Service ID had within that fingerprint (see Figure \ref{fig:scale_fingerprints_invocations}). Overall, this suggests that a major contributor to Service ID inflation is functionality that appears under different Service IDs across the dataset despite likely representing the same front-end functionality. Therefore, from hereafter we use the Service ID groupings based on these \textit{call fingerprints} rather than raw Service ID values, which we argue better represent the true repertoire of the front-end functionality provided by Alibaba's deployment.

\begin{figure}[h]
  \centering
  \includegraphics[width=0.99\columnwidth]{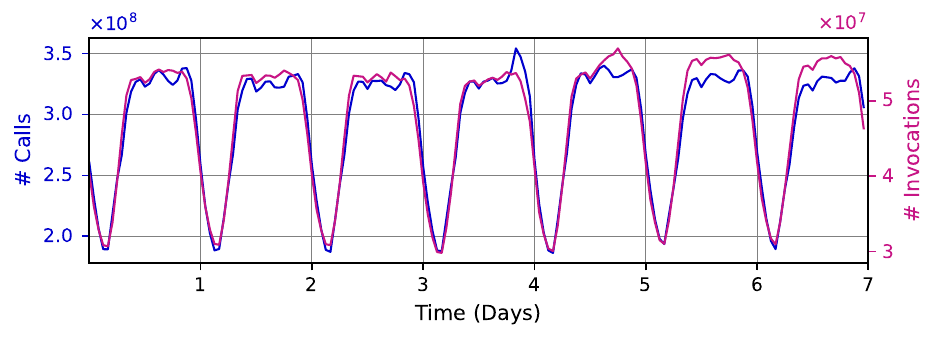}
  \caption{The total number of MS calls (blue) and front-end service invocations (pink) per hour across 7 days.}
  \label{fig:scale_workload}
\end{figure}

\subsection{Workload}
\label{sec:scale_workload}

\textbf{The total number of front-end requests and MS calls is measured in billions:} overall, there were 97 billion calls made between MSs associated with 15 billion front-end service invocations, reflecting the recruitment of multiple MSs to provide each front-end functionality.

\textbf{Workload significantly fluctuates across each day:} the number of calls and invocations fluctuated significantly across the 14 days, with a strong 24-hour periodicity, see Figure \ref{fig:scale_workload}. The average daily trough-to-peak increase was 80\% for MS calls and 79\% for front-end invocations. The considerable difference between lowest and highest total daily workload highlights the importance of predictive scaling, with inaccurate scaling for peaks and troughs potentially leading to wasted resource usage or impacted user experience respectively.

\subsection{Microservice Dependency Topology}
\label{sec:scale_dependencies}

\textbf{MSs are sparsely interconnected:} we constructed the MS dependency network from all calls made between MSs across the 14 days of operation, with a directed edge between nodes indicating at least one call was made between those MSs over the 14 days. Overall, we found 543,948 edges between MS, which is significantly lower than the theoretical maximum of a fully connected network of the same size (4.2 Billion), and is considerably sparser than reported by Meta (0.0001 vs 0.001). The 10-fold higher density of edges found at Meta may be a by-product of the MS granularity conjectured in Section \ref{sec:scale_ms_msi}, if MSs at Meta encapsulate broader functionality, each MS may have more interactions with other MSs to fulfil this functionality.

\subsection{Scale Implications}
\label{sec:scale_implications}

\noindent\textbf{The number of front-end service functionalities and back-end MSs within an MSA can be extremely large:} the number of MSs in the dataset (Section \ref{sec:scale_ms_msi}) is several orders of magnitude larger than popular testbeds \cite{gan_open-source_2019, noauthor_fudanselabtrain-ticket_2025, noauthor_googlecloudplatformmicroservices-demo_2025, noauthor_ocp-power-demossock-shop-demo_2025}. This disparity in scale likely contributes to management approaches assuming moderate to small deployment sizes. For example, one set of popular approaches construct causal graphs to capture the relationships between MS metrics during a fault \cite{ma_automap_2020, chen_causeinfer_2019, qiu_causality_2020, wang_cloudranger_2018, mariani_localizing_2018, wu_microdiag_2021, ma_ms-rank_2019, ma_servicerank_2022, lin_facgraph_2018, xin_causalrca_2023}. However, the causal structural modelling methods used by these approaches have poor time complexity as the number of variables increases \cite{spirtes_causation_1993, colombo_order-independent_2014, shimizu_linear_2006, shimizu_directlingam_2011}. The implicit reliance on this assumption challenges the suitability of many prior approaches to real-world large-scale systems.

This disparity in deployment size also leads to another common assumption that an adequately representative labelled corpus can be collected through fault injection or historical root cause information \cite{zhou_latent_2019, kohyarnejadfard_system_2019, scheinert_learning_2021}. However, with such a large number of MSs and diverse set of front-end functionality they support, the feasibility of collecting enough labelled data to adequately cover the large combination of failure conditions is unclear even when leveraging only partial supervised learning \cite{zhang_putracead_2022, raeiszadeh_real-time_2023}. 

A strategy for reducing challenges associated with characterising the normal behaviour of a MSA is to model the behaviour of individual MSs or MSs associated with individual front-end services based on the assumption that per-front-end service modelling is more tractable \cite{liu_unsupervised_2020, yang_transparently_2021}. However, in this dataset, we identified 8,000 front-end services (Section \ref{sec:scale_services}), therefore training and maintaining a model for each individual front-end service may be costly. Furthermore, we identified a potential 1.2 million distinct front-end functionalities associated with these front-end services. As each front-end functionality is likely associated with its own usage patterns and call graphs, this challenges whether building a model for each front-end service substantially reduces the problem of scale and diversity. On the other hand, building individual models for all 1.2 million front-end functionalities is unlikely to be feasible outside of very lightweight approaches \cite{xu_lightweight_2017}.

\noindent\textbf{The workload experienced by the system fluctuates significantly across the day:} widely used testbeds generally use stationary workload generators \cite{tene_giltenewrk2_2024} to simulate user behaviour. This means that generated datasets only represent system behaviour under stationary conditions \cite{pham_baro_2024, li_dejavu_2022, lee_eadro_2023, zhang_deeptralog_2022}. Subsequently, management approaches being developed and benchmarked using these deployments implicitly assume that workload, metrics, and call graphs are stationary. Our results (Section \ref{sec:scale_workload}) reveal strongly fluctuating user demand, indicating that methods trained and evaluated under stationary conditions may not generalise well to real-world variable workloads. This is particularly true for approaches that leverage change point detection for anomaly alerting \cite{pham_baro_2024}.

\begin{figure}[h]
\centering
\subcaptionbox{CDF of instances per microservice (Alibaba).%
\label{fig:heterogeneity_instances}}[0.55\columnwidth]{%
  \includegraphics[width=\linewidth]{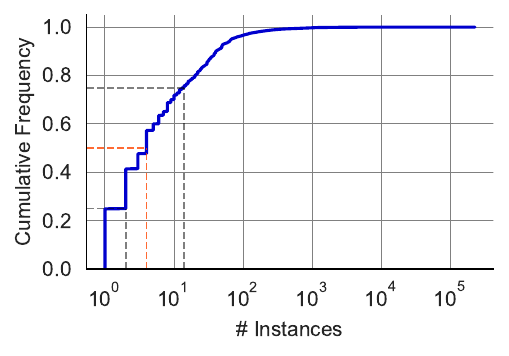}
}
\hspace{0.02\columnwidth}
\subcaptionbox{Per-MS instance counts comparison.%
\label{tab:instances}}[0.4\columnwidth]{%
  \footnotesize
  \setlength{\tabcolsep}{3.5pt}\renewcommand{\arraystretch}{1.05}
  \begin{tabular}{@{}lSS@{}}
    \toprule
     & {Alibaba} & {Meta} \\
    \midrule
    Min     & 1      & 1 \\
    Median  & 4      & 34 \\
    Average & 28     & 668 \\
    P99     & 309    & 9220 \\
    Max     & 227049 & 557156 \\
    \bottomrule
  \end{tabular}
}
\caption{Alibaba’s instance distribution and comparison with Meta:  
   (a) CDF of instances per microservice with with lower and upper quartiles (grey dashed) and median (red dashed);  
   (b) summary statistics highlighting the contrasting instance counts.}
\label{fig:cdf_plus_table}
\end{figure}

\section{Heterogeneity}
\label{sec:intro_heterogeneity}

MSAs are lauded for their flexibility and scalability \cite{wolff_microservices_2016, newman_building_2021}, achieved through independently developing and deploying loosely coupled MSs. Yet, this independence leads to a diverse and complex ecosystem. In practice, MSA heterogeneity manifests in many dimensions, from the distribution of instances across hosts and the co-location of multiple MSs in shared hardware, to distribution of workloads and the intricate web of dependencies that link them together. Furthermore, MSs can be variably integrated to provide diverse front-end service functionality, generating further avenues for heterogeneity. In this section, we explore how these facets of heterogeneity manifest in this dataset.

\textbf{The number of deployed instances per MS is highly imbalanced:} as shown in Figure \ref{fig:heterogeneity_instances}, the distribution of the number of instances per MS is heavy-tailed. 24.8\% of MSs were deployed as a single instance, with the median number of instances being 4. On the other hand, the highest number of instances was over 227,000, making up for 12.7\% of all deployed MS instances across the 14 days. Furthermore, the top 1\% of MSs accounted for around 895,000 instances, making up half of all deployed MS instances. Since we aggregated the number of deployed instances across the full dataset, the number of instances can either mean that a MS is generally deployed with a large number of instances or that a MS is regularly horizontally scaled. However, this dataset includes MSs from core applications that undergo much less auto-scaling (see \ref{sec:dataset}), this can be observed in our findings in Section \ref{sec:dynamicity_scaling}.


\textbf{MSs at Alibaba and Meta demonstrate similar distributions of, and maximum, instance counts:} the deployed instances per MS at both MSAa demonstrate a heavy-tail (see Figure \ref{tab:instances}), however, the median, average and 99-th percentile values for Meta are an order of magnitude higher. Given that the number of instances deployed at Meta is also an order of magnitude higher (see Section \ref{sec:scale_ms_msi}), this means that the central tendencies for both distributions when normalising by the total number of instances are similar, and thus the typical MS within both deployments represents a similar fraction of the overall deployed instances. Notably, we found that the maximum number of instances for a single MS were remarkably similar (227,000 vs 557,000) despite this order of magnitude difference in the total number of instances. This indicates that the heavy tail is more pronounced in this dataset, with the MS with the most instances accounting for 12.4\% of the total number of instances compared to Meta's 4.6\%. The convergence of maximum values despite the difference in the total number of instances might suggest that there are common operational ceilings when it comes to scaling. This ceiling could be a result of coordination overhead, such as managing load balancing between a larger number of instances, or non-linear performance gains due to resource contention for shared resources such as databases, caches, and network bandwidth.

\begin{figure*}[h]
  \centering
  \begin{subfigure}[t]{0.32\textwidth}
    \centering
    \includegraphics[width=\linewidth]{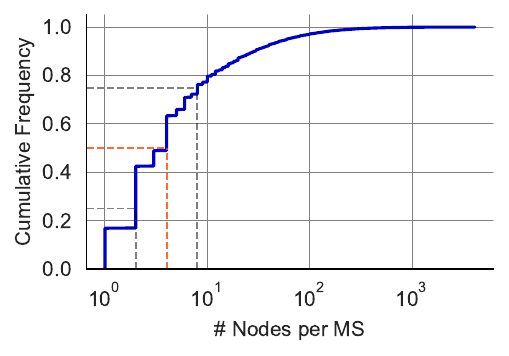}
    \caption{Nodes per MS}
    \label{fig:heterogeneity_nodes_ms}
  \end{subfigure}\hfill
  \begin{subfigure}[t]{0.32\textwidth}
    \centering
    \includegraphics[width=\linewidth]{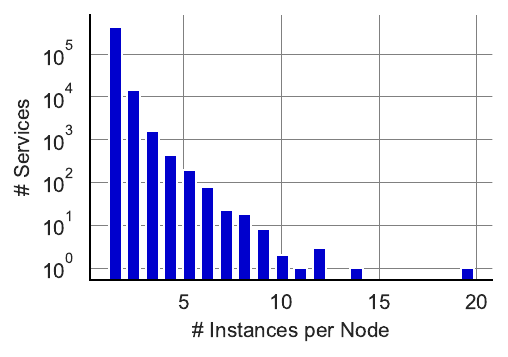}
    \caption{Max instances per node}
    \label{fig:heterogeneity_instances_node}
  \end{subfigure}\hfill
  \begin{subfigure}[t]{0.32\textwidth}
    \centering
    \includegraphics[width=\linewidth]{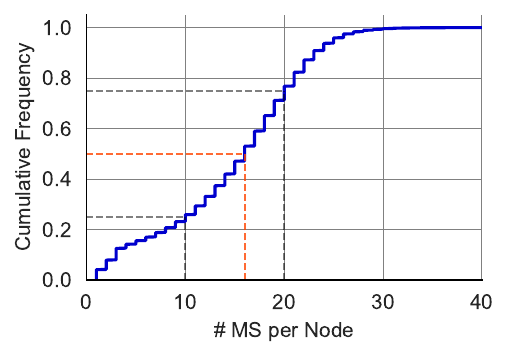}
    \caption{Unique MS per node}
    \label{fig:heterogeneity_ms_node}
  \end{subfigure}
  \caption{Deployment of MSs across nodes: (a) the number of nodes each MS is deployed across, (b) the maximum number of instances of the same MS per node, (c) the number of unique MSs per node, with upper and lower quartiles (grey dashed) and median (red dashed).}
  \label{fig:heterogeneity_top_threepanel_2col}
\end{figure*}

\subsection{Nodes and Microservices}
\label{sec:heterogeneity_nodes_ms}




\textbf{MS instances can be deployed across a large number of host machines:} as MS instances may be transient, we analysed the MS instances and their associated nodes within the first minute of the \textit{resource data}. This time period was arbitrarily chosen, but assumed to be representative of the dataset. Our analysis revealed that 84\% of these MSs were concurrently deployed on more than one node - with the median being 4 nodes and one MS spanning as many as 4,114 nodes (see Figure \ref{fig:heterogeneity_nodes_ms}). Moreover, we found that 96\% of MSs had a single instance per node, as shown in Figure \ref{fig:heterogeneity_instances_node}. These findings are a result of a deliberate orchestration strategy to aid load balancing and reliability by spreading MSs across nodes (see \ref{sec:dataset}). As found in our results, in a minority of cases, MSs may also be deployed multiple times on the same node to reduce fragmentation (see \ref{sec:dataset}).


\textbf{MS instances generally cohabit with a large number of other MS:} we found that nodes hosted between 1 and 38 MS instances, on average housing 15 MS instances each, with only 9.1\% of nodes hosting a single MS instance (see Figure \ref{fig:heterogeneity_nodes_ms}). This demonstrates the widespread use of cohabitation to minimise resource underutilisation.

\begin{figure}[h]
  \centering
  \includegraphics[width=0.55\columnwidth]{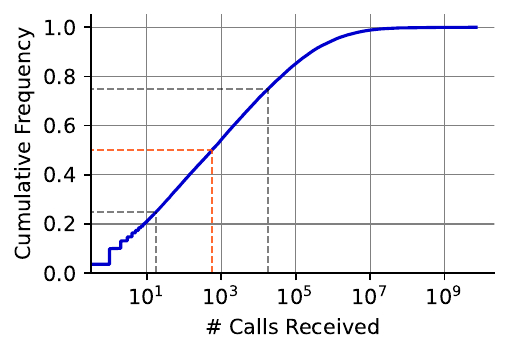}
  \caption{Cumulative frequency of the number of calls received per MS, with lower and upper quartiles (grey dashed) and median (red dashed).}
  \label{fig:heterogeneity_workload}
\end{figure}

\subsection{Microservice Workload}
\label{sec:heterogeneity_workload}

\textbf{The distribution of workload across MSs is extremely uneven:} the accumulation of calls from upstream MSs over a period of time can be defined as MS workload \cite{luo_power_2022}. To investigate how this workload was distributed across MSs, we aggregated the total calls received by each MS over the full 14-day period. Our analysis revealed a vast variation in workload across MSs, receiving from a single call to as many as 6.9 billion calls, the latter representing 7.1\% of all calls made over the two weeks (see Figure \ref{fig:heterogeneity_workload}). The cumulative workload distribution was heavy-tailed, with 21\% of MSs receiving fewer than 10 calls, whereas the top 1\% accounted for 87\% of all calls. These workloads are likely to be associated with e-commerce application traffic, such as Taobao, which generates the most workload for the Alibaba deployment (see \ref{sec:dataset}). These results corroborate previous research reporting heavy-tailed MS CPU workloads at Alibaba \cite{luo_power_2022}.



\begin{figure}[t]
\centering
\subcaptionbox{CDF of aggregate in- and out-degree per MS (Alibaba).\label{fig:heterogeneity_in_degree}}[0.55\columnwidth]{%
  \includegraphics[width=\linewidth]{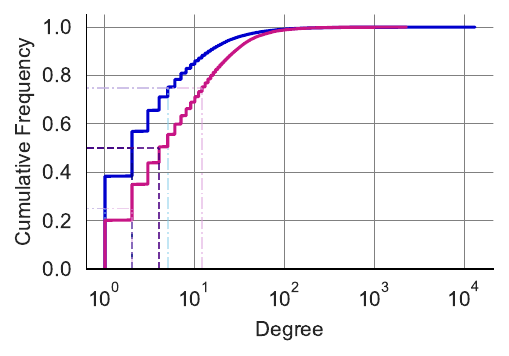}
}
\hspace{0.05\columnwidth}
\subcaptionbox{Degree comparison between Alibaba and Meta.\label{tab:degrees}}[0.65\columnwidth]{%
  \footnotesize
  \setlength{\tabcolsep}{3pt}\renewcommand{\arraystretch}{1.05}
  \begin{tabular}{@{}lcc|cc@{}}
    \toprule
         & \multicolumn{2}{c}{Alibaba} & \multicolumn{2}{c}{Meta} \\
    \cmidrule(lr){2-3}\cmidrule(lr){4-5}
         & In-Degree & Out-Degree & In-Degree & Out-Degree \\
    \midrule
    Min        & 1      & 1      & 1      & 1      \\
    Median     & 2      & 4      & 4      & 4      \\
    Average    & 8.2    & 12.2   & 14     & 12     \\
    P99        & 89     & 110    & 86     & 101    \\
    Max        & 13{,}030 & 2{,}245 & 14{,}084 & 5{,}865 \\
    \bottomrule
  \end{tabular}
}

\caption{Degree distribution and Alibaba–Meta comparison:
(a) CDF of aggregate in-degree (blue) and out-degree (pink) per microservice at Alibaba,
with lower/upper quartiles (dash–dot) and median (dashed);
(b) summary statistics comparing in- and out-degrees across Alibaba and Meta.}
\label{fig:degree_plus_table}
\end{figure}

\subsection{Microservice Dependencies}
\label{sec:heterogeneity_dependencies}

A complementary perspective to the MSA's global dependency structure explored in Section \ref{sec:scale_dependencies}, comes from analysing the local structure, such as MS in-degree and out-degree \cite{luo_characterizing_2021,huye_lifting_2023}. Here, the in-degree of a MS indicates how many other MSs call it, while the out-degree reflects how many other MSs it calls. 
We argue that these more aptly capture the importance of a MS within the architecture. A high in-degree signals a critical hub that many MSs rely on and is shared across a large amount of front-end functionality, and a high out-degree suggests a highly composite MS that aggregates the functionality of many downstream MSs.

\textbf{A small number of MSs have a disproportionately large number of dependencies:} using the aggregate dependency network constructed in Section \ref{sec:scale_dependencies}, we found that both in-degree and out-degree distributions exhibited a heavy-tail (see Figure \ref{fig:heterogeneity_in_degree}). Of the MSs, 38.5\% had an in-degree of 1, with a median of 2, yet the maximum in-degree reached 13,030, with one MS being called by 19.2\% of all other MSs. In contrast, only 20.2\% of MSs had an out-degree of 1, and the median out-degree was 4. However, the maximum out-degree was lower at 2,245 - meaning that 3.31\% of all MSs are called by this MS. Despite the higher maximum value for in-degree, these distributions indicate that generally most MSs call more peers than they are called by, in contrast to the findings of Meta \cite{huye_lifting_2023}, but supported by previous research at Alibaba \cite{luo_characterizing_2021}. Similarly to workload, many MSs with high numbers of dependencies are likely associated with e-commerce applications (see \ref{sec:dataset}. However, it is also likely that some also represent common core MSs that are shared between applications and business domains \cite{newman_building_2021} which may be central to overall system functionality; these architectural hot spots have also been identified in previous research \cite{luo_characterizing_2021}.

\textbf{Alibaba and Meta's local dependency topology structure is similar:} in Section \ref{sec:scale_dependencies} we noted that this dataset's global dependency structure was considerably sparser than that of Meta's. Yet, we found the local degree distributions to be strikingly similar, as shown in Table \ref{tab:degrees}. Greater sparsity, in conjunction with a larger number of MSs, and a local structure resembling Meta's, indicates that new MSs are incorporated by establishing dependencies with only a limited subset of peers. 

\begin{figure*}[h]
  \centering
  \begin{subfigure}[b]{0.4\textwidth}
    \includegraphics[width=\linewidth]{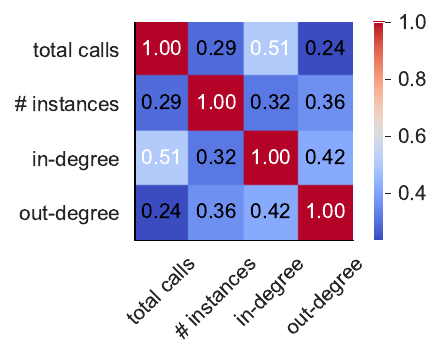}
    \caption{Kendall’s Tau correlation}
    \label{fig:heterogeneity_correlation}
  \end{subfigure}
  \hspace{0.1\textwidth}
  \begin{subfigure}[b]{0.4\textwidth}
    \includegraphics[width=\linewidth]{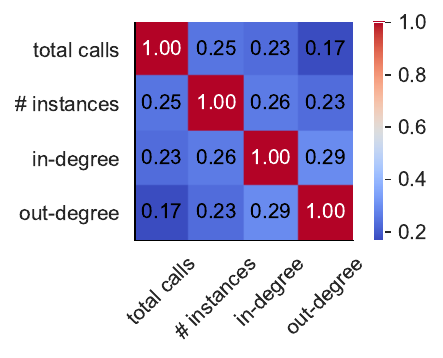}
    \caption{Jaccard similarity for top 1\%}
    \label{fig:heterogeneity_top_jaccard}
  \end{subfigure}

  \caption{(a) Correlation matrix of the Kendall’s Tau coefficient and (b) Jaccard similarity for the top 1\% across MS total calls received, number of instances, aggregate in-degree, and aggregate out-degree.}
  \label{fig:heterogeneity_map_both}
\end{figure*}

This observation, coupled with the similarity in local dependency structures, that likely arise from different organisational structures, technology stacks, and scales at Alibaba and Meta, suggests a convergence in MSA design. Such convergence likely arises due to universal constraints, such as performance, reliability, and maintainability, driving MSA design toward bounded complexity.

\subsection{Instances, Workload, and Dependencies}
\label{sec:heterogeneity_relationships}

\textbf{There is only a weak relationship between the number of MS instances and dependencies, and their workload:} so far we demonstrated that instances, workload, and dependencies all display heavy-tailed distributions. 
Therefore, we explore whether there exists a relationship between these metrics using Kendall's Tau \cite{kendall_1938}, a rank-based correlation measure that is less sensitive to extreme values than the conventional Pearson correlation coefficient. The resultant correlation matrix shown in Figure \ref{fig:heterogeneity_correlation} revealed a weak relationship between all studied variables. The highest correlation was found between total calls received and in-degree, suggesting that, to some extent, MSs that receive a wider array of different calls from more MSs also receive more calls overall; this provides some supporting evidence for the hub-like properties of some of these MSs discussed in Section \ref{sec:heterogeneity_dependencies}. We only found a weak relationship between total calls received and number of instances deployed for a specific MS. This indicates that other factors such as workload variability, functionality CPU requirements, and MS-specific scaling rules, play an important role in shaping the number of deployed instances of a MS.

\textbf{The top 1\% of MS with the most workload, instances, and number of dependencies are generally different:} 
we narrowed our investigation to the top 1\% of MSs for each category, to specifically understand whether these were generally the same outlier MSs. We calculated the Jaccard similarity between the set of MSs that comprised the top 1\% of each category to give us a similarity matrix, shown in Figure \ref{fig:heterogeneity_top_jaccard}. Our findings revealed that generally there is a lack of overlap between MSs in each set, with the lack of overlap being particularly pronounced for out-degree and total received calls, meaning that generally it is not the same set of MSs overrepresented in each category.

\begin{figure}[h]
  \centering
  \includegraphics[width=0.5\columnwidth]{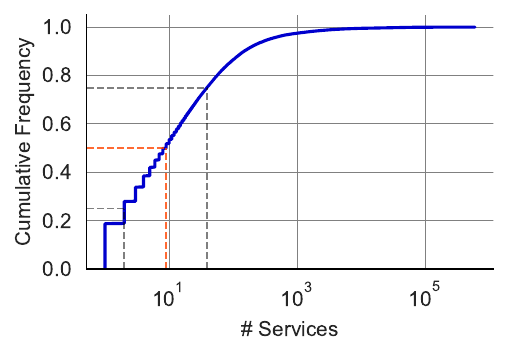}
  \caption{Cumulative frequency of the number of unique service \textit{call fingerprints} per MS, with lower and upper quartiles (grey dashed) and median (red dashed).}
  \label{fig:heterogeneity_ms_services}
\end{figure}

\subsection{Services and Microservices}
\label{sec:heterogneity_services_microservices}

\textbf{MSs can be shared across many front-end functionalities:} large-scale MSAs support many different applications which can cut across different business domains. Since each MS encapsulates a single functionality behind an interface, they are loosely coupled and composable, so varied front‑end requests can invoke the same MS. To understand how extensively each MS is used by different front-end functionalities, we leveraged the Service ID \textit{call fingerprints} from Section \ref{sec:scale_services}. For every MS in the dataset, we searched all \textit{call fingerprints} and recorded each that included that MS within its calls. Our results, Figure \ref{fig:heterogeneity_ms_services}, reveal that overall each MS was involved in a significant number of front-end functionalities, being associated with a median number of 9 and an average of 306 \textit{call fingerprints}. The distribution was heavy-tailed with the maximum number of front-end functionalities fulfilled by a single MS being 590,255, meaning that 49.8\% of all \textit{call fingerprints} involve this MS.

\begin{figure*}[t]
  \centering
  \begin{subfigure}[t]{0.32\textwidth}
    \centering
    \includegraphics[width=\linewidth]{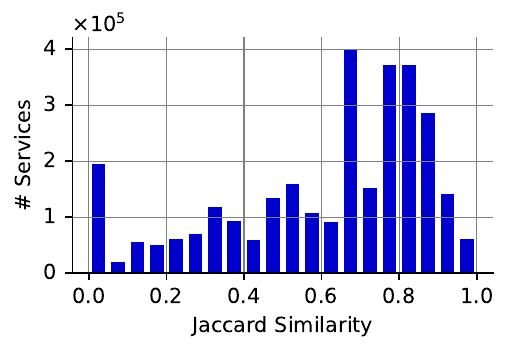}
    \caption{Jaccard Similarity}
    \label{fig:heterogeneity_jaccard}
  \end{subfigure}\hfill
  \begin{subfigure}[t]{0.32\textwidth}
    \centering
    \includegraphics[width=\linewidth]{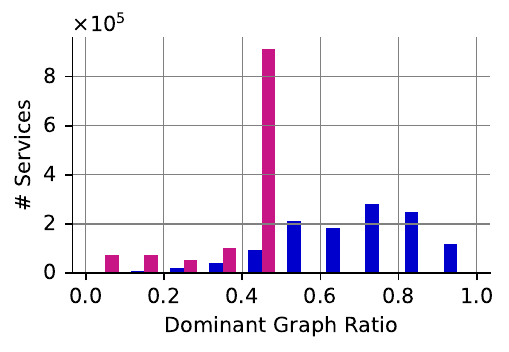}
    \caption{Dominant call ratio}
    \label{fig:heterogeneity_call_graph_dominance}
  \end{subfigure}\hfill
  \begin{subfigure}[t]{0.32\textwidth}
    \centering
    \includegraphics[width=\linewidth]{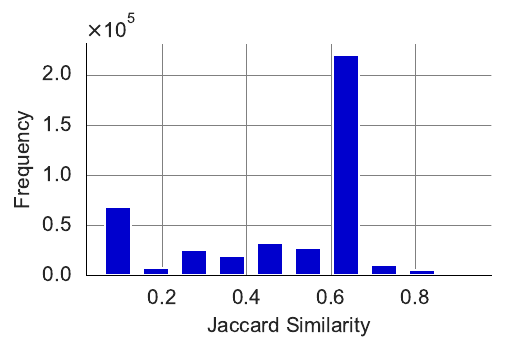}
    \caption{Dominant vs. non-dominant Jaccard}
    \label{fig:heterogeneity_dominant_similarity}
  \end{subfigure}
  \caption{Call graph variability analyses across services: (a) average Jaccard Similarity of call graphs per service (b) dominant call-graph ratio (blue) vs.\ uniform expectation (pink), and (c) Average Jaccard Similarity between dominant and non-dominant call graphs for highly dominant services.}
  \label{fig:heterogeneity_top_threepanel}
\end{figure*}

\subsection{Call Graph Variability}
\label{sec:heterogeneity_call_graph_variability}

To efficiently evaluate and compare the diversity of call graphs within the Alibaba dataset we represented traces as vectors, a common approach for embedding traces \cite{chen_tracemesh_2024}. These vectors are then converted into lower-dimensional representations using locality-sensitive hashing via MinHash \cite{broder_minhash_1997, indyk_lsh_1998}. The pairwise similarity of the resulting hash signatures approximates the Jaccard similarity \cite{broder_minhash_1997, indyk_lsh_1998, jaccard_similarity_1901}, quantifying how similar traces are in terms of the types of calls they contain. Whilst this approach does not directly compare graph structure we argue that the invocation order of MSs is not random, being based on underlying logic, and therefore two call graphs that share identical MS invocations are likely to be very similar in topological structure. Nevertheless, given this caveat, the results presented below may underestimate the true variability of call graphs but still provide a reasonable approximation.

A key challenge in quantifying the variability of call graphs is that inconsistencies in the observability data can impact results. For example, call graphs can be incomplete due to randomly missing data, meaning that two call graphs could be determined to be different simply due to missing data in one or both call graphs, rather than due to a true difference in the actual call patterns. Previous work has suggested that missing data is prevalent within the \textit{call data} released by Alibaba. However, this missing data is due to a systematic lack of observability for specific MSs within the deployment, and types of calls not published in the trace data (see \ref{sec:dataset}). Indeed, in a preliminary study of the first 12 hours of the dataset we found that only 0.12\% of traces with missing data could not be accounted for by systematic observability loss. Therefore, because the vast majority of missing data is deterministic, it is unlikely to significantly bias our vector-based comparisons of call graphs.

\textbf{The number of unique call graphs is measured in millions and front-end functionalities can be associated with a large number of different call graphs:} in total, we identified 79,737,924 unique call graph vectors. However, the median number of unique call graph vectors per service was only 1, with 59\% of Service IDs being associated with a single call graph vector, which is likely a product of the sporadic non-user facing functionality identified in Section \ref{sec:scale_services}. Nonetheless, a significant portion of Service IDs showed variability in their call graphs, with a mean number of unique call graph vectors per service of 31, and a maximum number of unique call graph vectors of just over 3 million. These values are higher than found in previous work at Alibaba \cite{luo_characterizing_2021}, which may be explained by the use of graph learning and clustering, grouping together graphs that demonstrate similarity in their call graph structure.

\textbf{Call graphs from the same service are not necessarily similar in their MS calls:} a large number of unique call graph vectors does not necessarily imply a high degree of call graph variability per service, as each call graph may only display small deviations in call patterns. Therefore, we investigated the average Jaccard similarity between call graph vectors for each front-end service that produced more than a single unique call graph vector. As shown in  Figure \ref{fig:heterogeneity_jaccard}, the distribution of similarities, considering the frequency of each call graph occurrence shows a grouping of high average similarities between 0.65 and 0.9, accounting for 52.8\% of all services. On the other hand, we also found that a substantial number of services (28.5\%) demonstrated less than 0.5 similarity between their call graph vectors on average, with 6.5\% of service showing no similarity. As we account for the number of times each call graph vector occurred in our analysis, the cluster of higher average similarities may be caused by dominant call graph vectors per service. This follows from design principles, where a given front-end feature typically relies on a core set of MS calls with small deviations occurring due to conditional or alternative functionality.




\textbf{Front-end functionality often has a dominant call graph:} to further understand the prevalence of dominant call graphs, we explored the proportion of the total call graph vectors produced by each service that matched the most commonly occurring call graph. In a first analysis we found that there were a large number of services (13.4\%) that only produced two unique call graphs each only appearing once, resulting in a spike of ratios at 0.5, these cases do not provide much information as to the prevalence of dominant call graphs and thus we excluded them from the subsequent analysis. The resulting distribution, shown in Figure \ref{fig:heterogeneity_call_graph_dominance}, reveals a significant skew of the ratio of the dominant call graph vector towards higher values than would be expected if all call graph vectors had an equally likely probability to occurrence, with 82\% of Service IDs demonstrating higher ratios than the maximum possible ratio if all call graphs were equally likely. This confirms that services generally have a dominant call graph that disproportionally occurs during service invocation, with rarer non-dominant call graphs.

\textbf{Non-dominant do not necessarily show high similarity to the dominant call graph pattern of a front-end service's functionality:} subsequently, we sought to characterise the level of similarity between dominant and non-dominant call graphs. We focused our investigation on services with a high prevalence of dominant call graphs ($>$0.75), for each Service ID we found the average Jaccard similarity between the dominant call graph vector and all other call graph vectors. The resulting distribution, shown in Figure \ref{fig:heterogeneity_dominant_similarity}, reveals that non-dominant call graphs show some similarity to non-dominant call graphs, with an average Jaccard similarity of 0.6. However, we found that 35\% of non-dominant call graphs share less than 50\% of the same calls, with 15\% of non-dominant call graphs demonstrating zero similarity with corresponding dominant call graphs. Our findings extend previous observations of variability in MS call graphs within large-scale MSAs at both Alibaba and Meta \cite{luo_characterizing_2021, huye_lifting_2023}. We found that not only are call graphs from the same front-end functionality or service potentially highly variable, but non-dominant call graphs can show dissimilarity in the types of MS calls involved. We note that the total number of unique call graphs is very large (79 million) even when not accounting for call graph structure.

\subsection{Heterogeneity Implications}
\label{sec:heterogeneity_implications}

\noindent\textbf{MSs can be multiplexed by a large variety of other MSs and front-end functionality:} a common assumption employed by many fault management approaches is that the behaviour of MSs is naturally correlated with the usage of front-end applications and their KPIs, and thus deviations in this relationship can be used to detect anomalies \cite{samir_dla_2019, samir_dla_2019} and correlation with faulty KPIs can help localise the root cause of faults \cite{wang_cloudranger_2018, ma_ms-rank_2019, ma_servicerank_2022, li_practical_2021, liu_microhecl_2021}. However, we found that MSs can be involved with potentially thousands of front-end service functionalities (Section \ref{sec:heterogneity_services_microservices}), inevitably resulting in signal mixing as the behaviour of MSs is the aggregate of many different front-end usage patterns. Without appropriate treatment of signal mixing, for instance through blind source separation \cite{pal_blind_2013}, these approaches are likely to incorrectly detect and attribute faults.

Another assumption employed by anomaly detection approaches is that metrics are approximately normally distributed and well behaved, allowing the use of the 3-sigma rule 3-Sigma rule \cite{lin_microscope_2018, yu_microrank_2021, li_causal_2022, lee_eadro_2023} or the extreme studentized deviate test \cite{vallis_novel_2014} to highlight outlier values. Instead, our findings demonstrate that MSs can be multiplexed by many front-end workloads, potentially producing non-Gaussian mixtures with varying means and variances, which can yield unexpected tail behaviours and non-stationary signals as different workloads dominate over time. These characteristics challenge the suitability of these statistical approaches in real-world systems which would experience increased false alarms and decreased accuracy under such conditions.

Another set of popular approaches for leveraging distributed tracing data for management is to treat calls within traces as sequences of events \cite{nedelkoski_anomaly_2019, meng_midiag_2020, purfallah_mazraemolla_effective_2024}. Sequence-learning approaches, such as recurrent neural networks, can then be used to learn the normal sequences of calls between MSs. Trained models can then be used to flag anomalous calls that fall below some threshold of expected next events. However, these approaches rely on the assumption that the number of next possible calls is moderate. However, we found that MSs can have very large out-degrees (Section \ref{sec:heterogeneity_dependencies}), meaning a vast number of possible next events. This can dramatically increase the amount of training data required for such approaches. Additionally, a dispersed probability distribution over many potential next events could make approaches sensitive to threshold parameters. Furthermore, picking a one-size-fits-all threshold when the number of next events across MSs is heavy-tailed is likely to result in increased false negatives and increased false positives for MSs with low and high out-degree respectively.

\noindent\textbf{Large‑scale MSAs generate many call graphs, with each front‑end function following a common structure but sometimes producing rare, divergent ones:} an assumption of many management approaches is that normal call graphs are adequately represented within tracing data. This allows learning approaches to model the structure of normal call graphs \cite{yang_transparently_2021, panahandeh_serviceanomaly_2024, wang_workflow-aware_2020, meng_detecting_2021} which can be subsequently used for anomaly detection and localisation. Our analysis shows that while most front-end functions exhibit common call graph patterns, a substantial number of rare, divergent graphs occur (Section \ref{sec:heterogeneity_call_graph_variability}). Such rare graphs are easily under-represented, especially under typical trace-sampling policies, resulting in incomplete models and elevated false-positive rates at runtime.

Another assumption of management approaches is that the number of call graphs produced by a deployment is moderate, making such modelling tractable \cite{yang_transparently_2021, panahandeh_serviceanomaly_2024, wang_workflow-aware_2020, meng_detecting_2021}. In contrast, we observed 79 million unique call graphs and an average of 44 million call graphs per hour. This scope challenges the computational feasibility of approaches that rely on enumerating or clustering call graphs to provide real-time insights into production systems.

\noindent\textbf{The spread of workload across MSs is highly uneven:} many auto-scaling and resource-management approaches implicitly assume that most MSs experience substantial and predictable demand, justifying uniformly sophisticated workload-prediction models \cite{luo_power_2022, choi_phpa_2022, abdullah_burst_2022, nguyen_graph_2022, cheng_proscale_2023}. Our findings (Section \ref{sec:heterogeneity_workload}), consistent with prior observations \cite{luo_characterizing_2021, luo_power_2022}, reveal a strongly long-tailed distribution of workload, with a small fraction of MSs handling the bulk of requests. This suggests that the greatest gains in resource efficiency and service level agreement compliance will come from improving prediction for high-demand MSs, whereas the majority may be adequately handled by simpler heuristic or rule-based scaling. 

\noindent\textbf{MS instances are deployed across a large number of nodes and can be co-located with a large number of different MS instances:} many management approaches leverage distributed tracing data to capture the direct dependencies between MSs \cite{liu_microhecl_2021, yang_transparently_2021, panahandeh_serviceanomaly_2024, yu_microrank_2021, yu_tracerank_2023, gan_sage_2021, gan_sleuth_2024}, implicitly assuming that faults only propagate along these edges. Our analysis hows that MS instances are deployed across a large number of nodes and are often co-located with many other MSs (\ref{sec:heterogeneity_nodes_ms}). Such co-location enables indirect fault propagation through shared resources even in the absence of direct calls \cite{ma_automap_2020}. This finding challenges the adequacy of models that rely solely on direct call relationships for reasoning about fault propagation in large-scale MSAs.

\section{Dynamicity}
MSAs are inherently dynamic. Whilst dependencies are fixed in underlying code, at runtime, calls between MSs are transient, only occurring based on a front-end request. Furthermore, the deployment of MS instances is elastic, with instances being scaled up and down in response to changes in demand and performance. Furthermore, the architecture is prone to change as MSs are added, removed, and updated to evolve the functionality of the system \cite{newman_building_2021}. In this section, we explore how this dynamicity manifests in the dataset.

\begin{figure*}[h]
  \centering
  \begin{subfigure}[b]{0.45\textwidth}
    \includegraphics[width=\linewidth]{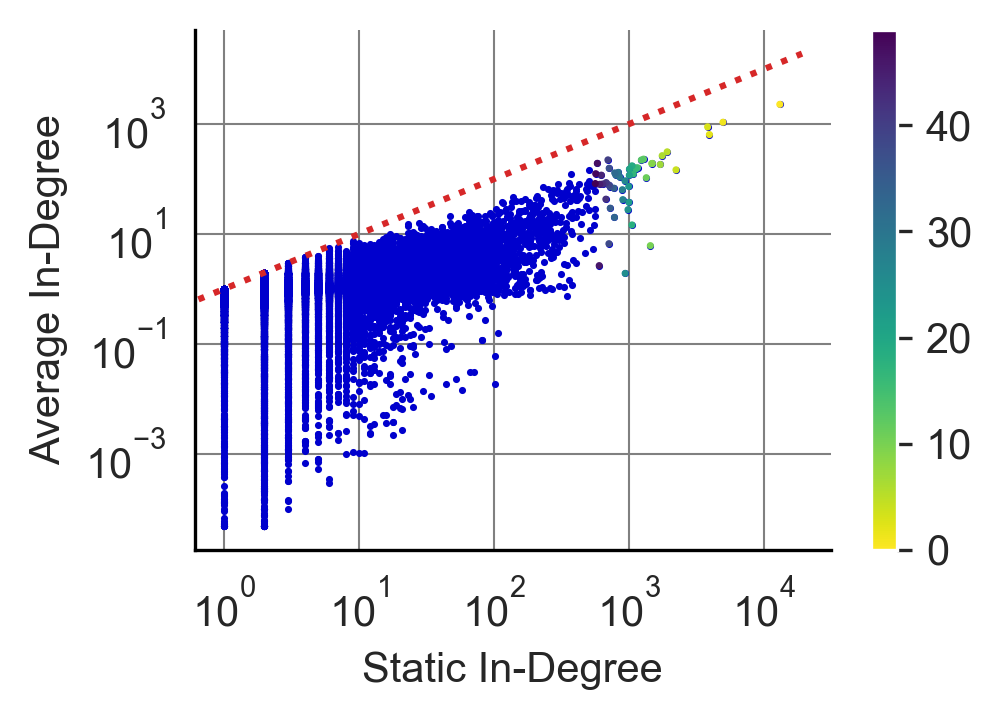}
    \caption{Temporal vs.\ static in-degree}
    \label{fig:dynamicity_in_scatter}
  \end{subfigure}
  \hspace{0.05\textwidth}
  \begin{subfigure}[b]{0.45\textwidth}
    \includegraphics[width=\linewidth]{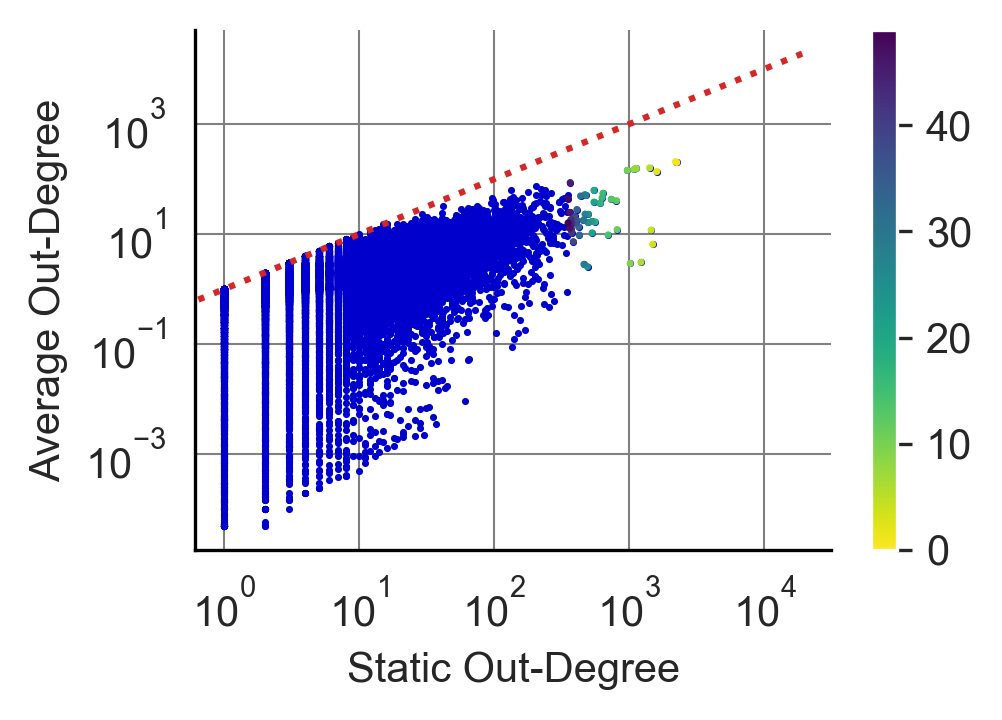}
    \caption{Temporal vs.\ static out-degree}
    \label{fig:dynamicity_out_scatter}
  \end{subfigure}\\[1ex]

  \begin{subfigure}[b]{0.45\textwidth}
    \includegraphics[width=\linewidth]{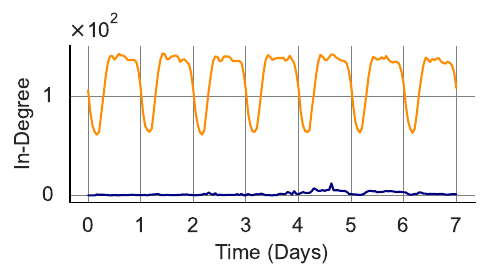}
    \caption{In-degree per hour for ranks 25 \& 26}
    \label{fig:dynamicity_in_example}
  \end{subfigure}
  \hspace{0.05\textwidth}
  \begin{subfigure}[b]{0.45\textwidth}
    \includegraphics[width=\linewidth]{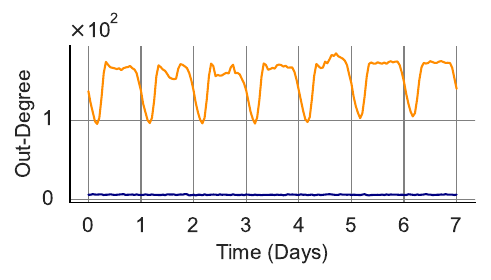}
    \caption{Out-degree per hour for ranks 4 \& 5}
    \label{fig:dynamicity_out_example}
  \end{subfigure}

  \caption{Temporal degree analysis: comparison of average temporal and static (a) in-degree and (b) out-degree with the line of equality (red dotted), and example time series of the temporal (c) in-degree for the 25\textsuperscript{th} (orange) and 26\textsuperscript{th} (blue) ranked static in-degree and (d) out-degree for the 4\textsuperscript{th} (orange) and 5\textsuperscript{th} (blue) ranked static out-degree.}
  \label{fig:dynamicity_all}
\end{figure*}

\subsection{Time-Varying Dependencies}
A common perspective when discussing MSs within a MSA is the overall dependency structure between MSs \cite{huye_lifting_2023, gluck_introducing_2020, gamage_using_2021}, as we have done in Section \ref{sec:scale_dependencies} and \ref{sec:heterogeneity_dependencies}. However, this view fails to account for differences in the behaviour of the system at runtime. Indeed, more recent research has shown the value of investigating the temporal behaviour of MSAs at runtime \cite{parker_visualizing_2023, winchester_temporal_2023}. Still, little research has been carried out to understand to what extent the temporal behaviour of a MSA differs from the aggregate view of the system. As a first step to contributing to this question, we explored the differences in time-varying degree of MSs at runtime by ingesting the \textit{call data} into the temporal graph database Raphtory \cite{steer_raphtory_2024}.

\textbf{Runtime dependencies can vary greatly from the static view of the system:}
\label{sec:dynamicity_temporal_dependencies}
we calculated the average temporal in- and out-degree for each MS, sampled in 1-minute snapshots, and compared these against their corresponding static degrees from Section \ref{sec:scale_dependencies}. Our results indicate that the temporal degree of MSs at runtime does not necessarily directly correspond to their degree in the static view of the system (see Figures \ref{fig:dynamicity_in_scatter} and \ref{fig:dynamicity_out_scatter}). For both in- and out-degree, we found that, whilst the highest static degree MSs tend to also have a high average degree, this relationship quickly deteriorates. For example, the 26\textsuperscript{th} highest static in-degree MS only has the 4,809\textsuperscript{th} highest average in-degree. This is even more apparent for out-degree, with the 5\textsuperscript{th} highest static out-degree MS only having the 19,774\textsuperscript{th} highest out-degree. 

To understand the differences between MSs at runtime that lead to these differences, we compared the runtime degree time series of the two outliers mentioned above with the MSs directly below them in static degree ranking, see Figure \ref{fig:dynamicity_out_example} and \ref{fig:dynamicity_in_example}. In both cases, we found that, when compared to the MS preceding it in the static degree ranking, these MS displayed much less regular changes in dependencies with weak daily periodicity and a significantly lower average number of dependencies on a one minute basis. This indicates that these MS are not involved in a large amount of system functionalities on a moment-by-moment basis, but rather call, or are called, by a large number of different MSs over time. This inflates their in- or out-degree in the static view of the system when compared to MSs that instead are consistently involved with a large number of other MSs on average at runtime.


\begin{figure}[h]
  \centering
  \begin{subfigure}[t]{0.5\columnwidth}
      \centering
      \includegraphics[width=\linewidth]{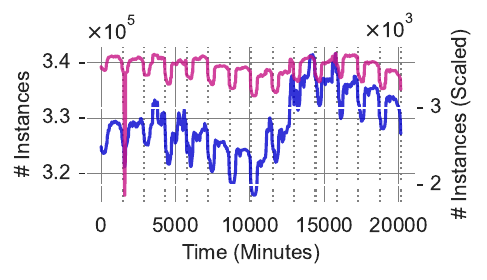}
      \caption{Number of deployed instances (blue) and instances normalized by each MS’s maximum number of instances (pink).}
      \label{fig:dynamicity_total_scaling}
  \end{subfigure}
  \hfill
  \begin{subfigure}[t]{0.4\columnwidth}
      \centering
      \includegraphics[width=\linewidth]{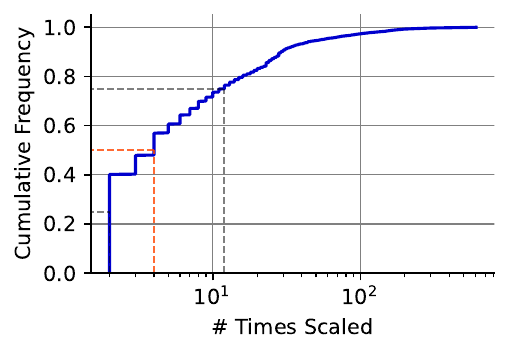}
      \caption{Cumulative frequency of the number of times MSs scaled over the 14 days, with lower and upper quartiles (grey dashed) and median (red).}
      \label{fig:dynamicity_scaling_dist}
  \end{subfigure}

  \caption{Dynamicity of MS deployments: (a) scaling patterns across instances, (b) distribution of scaling events.}
  \label{fig:dynamicity_combined}
\end{figure}

\subsection{Horizontal Scaling}
\label{sec:dynamicity_scaling}

The traces in this dataset only includes long running MSs which are a part of core applications that do not undergo much auto-scaling (see \ref{sec:dataset}). Nonetheless, to investigate the horizontal scaling within the deployment we used the \textit{resource data} which includes samples from all deployed MS instances every minute to create a time series, see Figure \ref{fig:dynamicity_total_scaling}. We found that 76.9\% of the 28,000 stateless MS maintained the same number of instances across the 14 days.



\textbf{MS that scale do not generally do so in response to daily fluctuations in demand:} of the remaining 33.1\% of MSs we found that only just over half of them (51.2\%) scaled more than twice over the 14 days (see Figure \ref{fig:dynamicity_scaling_dist}). To determine the number of MSs that scale daily we extracted both time and frequency domain features using the auto-correlation function (ACF) and fast Fourier transform (FFT). Using the ACF, we derived the ACF ratio by comparing the autocorrelation at a 24-hour lag to that at a lag of 1, providing a measure of the daily cycle's prominence relative to short-term correlations. Using the FFT derived power spectrum, we calculated the power ratio as a fraction of total spectral power concentrated at the 24-hour frequency. Using these two metrics of periodicity we identified 1,240 MS that demonstrated strong daily scaling, accounting for only 18.8\% of all MSs that scale (i.e., 4.4\% of all stateless MS).

\textbf{MS that scale daily can have a large change in the number of instances:}
For those MSs that demonstrated strong daily periodicity, we found that the average daily peak to trough change in the number of instances per MS was 13, meaning, on average, MSs experienced a 43.1\% change in their number of deployed instances per day.

\begin{figure}[h]
  \centering
  \includegraphics[width=0.95\columnwidth]{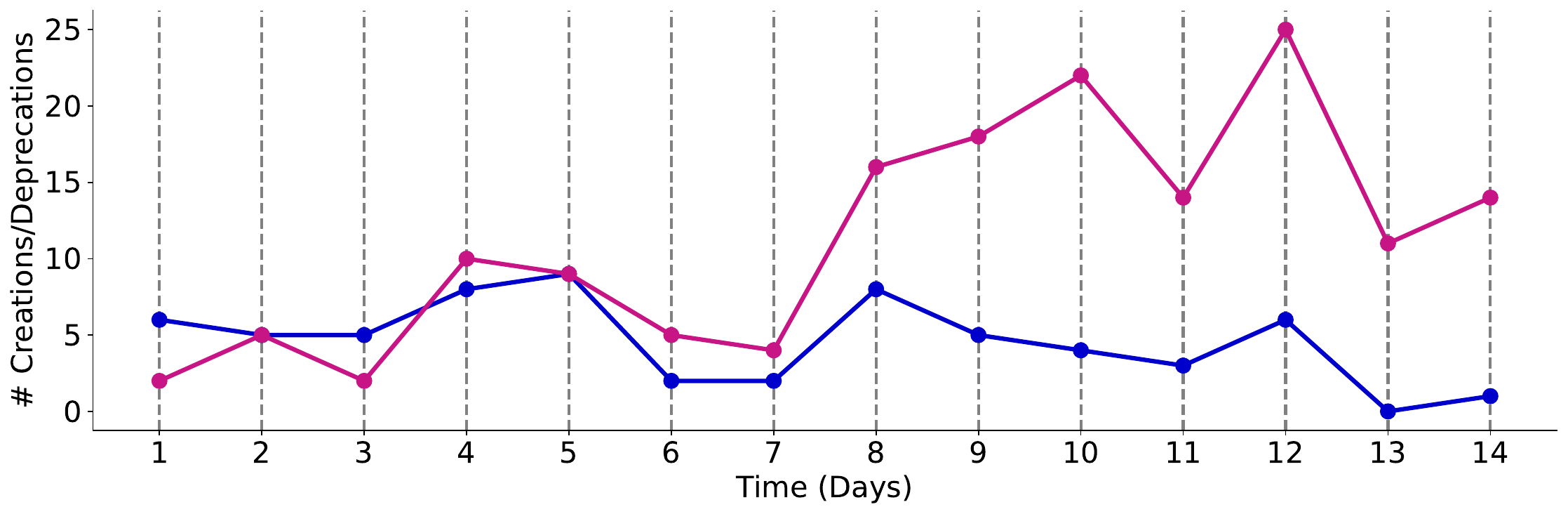}
  \caption{Number of MSs created (blue) or deprecated (pink) per day.}
  \label{fig:dynamicity_evolution}
\end{figure}

\subsection{Evolutionary Architecture}
\label{sec:dynamicity_evolutionary}

The decoupled nature of MSs provides bounded context, meaning that incremental changes can be made to individual MSs without breaking an application, leading to an architecture that grows and changes continuously \cite{newman_building_2021, huye_lifting_2023}. As information as to the creation and deprecation of MSs does not exist for this dataset, we assessed the extent to which the deployment undergoes such evolution using the \textit{resource data} due to its deterministic sampling of all deployed MS instances every minute.

\textbf{Churn occurs on a daily basis:} to differentiate between potential scale-to-zero or missing data and true deprecation and creation, we carried out two data processing steps. We first identified every MS with zero instances across the 14 days. Next, if a MS never had instances before or again they were flagged as creation or deprecation candidates respectively. If a candidate had more than one period of zero instances, we calculated the distribution of all period durations, and the pre-creation or post-deprecation period was compared to this distribution. Candidate periods which were greater than 3 standard deviations above the mean were considered genuine deprecation or creation events.

Results showed that across the 14 days, more MSs were deprecated than created, with 64 new MSs created and 157 MSs deprecated, see Figure \ref{fig:dynamicity_evolution}. Whilst the number of MS creations and deprecations are comparatively small compared to the size of the deployment, the number of new MSs added across two weeks alone was greater than the total size of most experimental testbeds (SocialNetwork - 26 MSs \cite{noauthor_deathstarbenchsocialnetwork_nodate}, MediaMicroservices - 30 MSs \cite{noauthor_deathstarbenchsocialnetwork_nodate}, TrainTicket - 68 MSs \cite{noauthor_fudanselabtrain-ticket_2025}, SockShop - 8 MSs \cite{noauthor_ocp-power-demossock-shop-demo_2025}, HipsterShop - 10 MSs \cite{noauthor_googlecloudplatformmicroservices-demo_2025}).

\textbf{Alibaba and Meta have similar rates of MS churn:}
When contrasting the number of MS creation and deprecation events within the dataset to those found by Meta, we found remarkably similar, albeit inverted, values with a median number of MS creations at 5 and 12, and the median number of deprecations at 10.5 and 4 per day for Meta and Alibaba, respectively. The higher deprecation values relative to creation for this dataset when compared to Meta's may suggest that this deployment, being significantly larger than that of Meta, is in a lifecycle of consolidation and removal of unnecessary MS bloat. However, it should be noted that Meta found significant jumps in creation and deprecation across their longer period of study \cite{huye_lifting_2023} and thus our findings across just 14 days may not accurately depict the broader trend, indeed up until day 6 creation and deprecation events are more in equilibrium. Notably, the median values found for both deployments are of similar magnitudes, despite a considerable difference in the total number of MSs (18,500 vs 28,000), suggesting that the rate of growth of large-scale MSAs may be independent of total size.

\begin{table}[htbp]
\centering
\caption{Common assumptions in MS management approaches, their validity in large-scale MSAs, and recommendations for future work.}
\begin{tabular}{p{0.28\linewidth} p{0.35\linewidth} p{0.37\linewidth}}
\toprule
\textbf{Assumption} & \textbf{Finding} & \textbf{Recommendation} \\
\midrule

Moderate deployment size. & 
Real deployments can comprise tens of thousands of MSs and millions of instances (\ref{sec:scale_ms_msi}). &
Design and evaluate methods for scalability using large-scale datasets/testbeds \cite{luo_power_2022, lee_tale_2024, detti_ubench_2023}.
 
\\
 
Availability of labelled data. & 
The enormous diversity of MSs and functionalities (\ref{sec:scale_ms_msi}, \ref{sec:scale_services}) makes creating an adequately representative labelled corpus impractical. &
Shift toward partial \cite{zhou_wsl_2017}, self- \cite{chen_simclr_2020}, or synthetic \cite{chawla_smote_2002} supervision.
 
\\
 
Per-service modelling is tractable. & 
Thousands of front-end services (\ref{sec:scale_services}) with distinct usage and call graphs makes per-service modelling challenging. &
Explore shared learning frameworks such as hierarchical or multi-task learning \cite{ruder_multitask_2017} to combine learning across services.
 
\\
 
Stationary workloads and call graphs. & 
System-level workload fluctuates across each day (\ref{sec:scale_workload}). &
Evaluate methods under realistic non-stationary workloads \cite{noauthor_locustio_nodate}.

\\
 
\midrule
Simple coupling between front-end KPIs to single MSs & 
MSs are multiplexed by thousands of front-end functionalities (\ref{sec:heterogneity_services_microservices}), likely causing signal mixing. &
Use latent-factor models \cite{jolliffe_pca_2011} or causal discovery with unobserved confounders \cite{spirtes_cps_2001}.

\\
 
Metric distributions are approximately normal. & 
Multiplexed workloads (\ref{sec:heterogneity_services_microservices}) likely produce non-Gaussian mixtures with changing means and variances. &
Replace Gaussian thresholds with robust heavy-tailed \cite{huber_robust_2009}, quantile-based \cite{koenker_quantile_2005}, or copula models \cite{nelsen_introduction_2006}.

\\
 
Number of potential next calls in a call graph is moderate. & 
Branching factor in traces can be very large (\ref{sec:heterogeneity_dependencies}). &
Adopt sub-graph pattern mining \cite{yan_gspan_2002} or attention-based sequence models \cite{vaswani_attention_2017}.

\\
 
Call graphs from the same front-end service are stable. & 
Rare, divergent call graphs are common (\ref{sec:heterogeneity_call_graph_variability})&
Apply graph-level embedding \cite{narayanan_graph2vec_2017} and clustering tolerant to structural variability \cite{vishwanathan_graph_2010}.

\\
 
Faults propagate only along explicit calls. & 
Heavy co-location of MS instances (\ref{sec:heterogeneity_nodes_ms}) creates indirect propagation paths. &
Integrate call-graph edges with shared-infrastructure links in propagation models \cite{boccaletti_multilayer_2014}

\\
 
All MSs merit equally sophisticated scaling policies. & 
Workload is highly skewed (\ref{sec:heterogeneity_workload}); most prediction benefit may lie in a small subset of MSs. &
Make resource-allocation and prediction policies total workload-aware.

\\
 
\midrule
Static dependency view is representative of the MSA. & 
Temporal interactions can differ greatly from static topology (\ref{sec:dynamicity_temporal_dependencies}). &
Combine temporal network representations \cite{masuda_guide_2016} with static views.

\\

MSA structure is stable enough that models trained once remain valid. & 
MSAs evolve constantly (\ref{sec:dynamicity_evolutionary}), altering the normal baseline. &
Explore continual \cite{parisi_continual_2019} and transfer learning \cite{pan_transfer_2010} frameworks for online adaptation.

\\

\bottomrule
\end{tabular}
\label{tab:assumptions_findings}
\end{table}

\subsection{Dynamicity Implications}
\label{sec:dynamicity_implications}

\noindent\textbf{The static view of system dependencies does not necessarily correspond to runtime interactions:} previous research has used the static view of dependencies to characterise MSAs \cite{huye_lifting_2023} including identifying design anti-patterns \cite{gamage_using_2021}, assuming that the static view accurately represents MS behaviour at runtime. However, our results (Section \ref{sec:dynamicity_temporal_dependencies}) show that temporal interactions among MSs can diverge substantially from this static view, with marked differences in temporal centrality. Consequently, applying graph measures to the static dependency graph \cite{gamage_using_2021} can yield characterisations that are misaligned with the system’s true runtime behaviour. Such analyses would benefit from complementing static representations with temporal network–based insights that capture runtime dynamics \cite{masuda_guide_2016}.


\noindent\textbf{MS architectures evolve on a daily basis:} of the commonly used MS testbeds \cite{gan_open-source_2019, noauthor_fudanselabtrain-ticket_2025, noauthor_googlecloudplatformmicroservices-demo_2025, noauthor_ocp-power-demossock-shop-demo_2025} only one \cite{noauthor_ciscodevnetbookinfo-cloudnative-sample_2025} incorporates versioning that changes the functionality of the deployment. This lack of evolving testbeds explains why most management approaches give limited attention to system evolution. A common general approach for management approaches is to model the ``normal'' behaviour of the system from historical data \cite{yu_microrank_2021,gulenko_detecting_2018,wu_microdiag_2021,wu_microrca_2020,jin_anomaly_2020,xu_lightweight_2017,hacid_performance_2021,scheinert_learning_2021} or tracing data \cite{wang_workflow-aware_2020,yang_transparently_2021,meng_detecting_2021,panahandeh_serviceanomaly_2024,nedelkoski_anomaly_2019,meng_midiag_2020,purfallah_mazraemolla_effective_2024}, implicitly assuming a largely stationary architecture. Our findings (Section \ref{sec:dynamicity_evolutionary}), consistent with Meta's observations \cite{huye_lifting_2023} and microservice design principles \cite{newman_building_2021}, show that MSAs change continuously: MSs and their dependencies appear, disappear, and are repurposed. Such changes inevitably alter call graph structure and, as MSs are multiplexed by new functionality, shift the expected patterns of their metrics. Although there are promising steps toward handling this non-stationarity, such as human-in-the-loop flagging of architectural changes \cite{purfallah_mazraemolla_effective_2024} and transfer learning \cite{gan_sleuth_2024}, the practical implications of system evolution on management approaches, including the overhead of frequent retraining, remain under-appreciated.

\section{Summary}

In Sections \ref{sec:scale_implications}, \ref{sec:heterogeneity_implications}, and \ref{sec:dynamicity_implications} we examined how our findings challenge common assumptions in MSA management approaches. Table \ref{tab:assumptions_findings} summarises these findings together with their implications and recommendations for future work. Overall, our results reveal that existing approaches systematically underestimate the complexity of real-world deployments, particularly with respect to (i) scalability (Section \ref{sec:scale_implications}), (ii) the fact that MSs may be multiplexed by many other MSs and front-end functionalities (Section \ref{sec:heterogeneity_implications}), (iii) the variability of call graphs from the same front-end functionality (Section \ref{sec:heterogeneity_implications}), (iv) the exposure of MSs to potential interference from co-habitation (Section \ref{sec:heterogeneity_implications}), and (v) adaptation to an evolving system (Section \ref{sec:dynamicity_implications}).

Many of the above limitations of current state-of-the-art approaches appear to stem, at least in part, from limited access to real-world MS deployments \cite{huye_lifting_2023}. The research testbeds that are commonly used throughout the literature \cite{gan_open-source_2019, noauthor_fudanselabtrain-ticket_2025, noauthor_googlecloudplatformmicroservices-demo_2025, noauthor_ocp-power-demossock-shop-demo_2025} are not sophisticated enough to capture the complexities of actual deployments, especially large-scale deployments such as at Alibaba and Meta \cite{huye_lifting_2023}. Therefore, we join recent calls for improved testbed realism \cite{seshagiri_sok_2022, huye_lifting_2023}. 

To bridge this gap, future testbeds should move beyond single-application deployments by supporting a broader range of front-end services (see Section \ref{sec:scale_services}). They should also 
capture the varying nature of real-world systems, such as enable architectural evolution through versioning \cite{detti_ubench_2023} and integrating workload generators with real-world service usage data that can generate service-specific demand fluctuations \cite{noauthor_locustio_nodate}. Such variability is crucial for exposing changes in dependency structure, including the appearance and disappearance of call paths over time (Sections \ref{sec:scale_workload} and \ref{sec:heterogeneity_call_graph_variability}). Finally, to reproduce the real-world variability of call graphs, testbeds should incorporate mechanisms known to alter call paths in practice, such as graceful degradation under load or failure, cache-driven eviction policies, and conditional user inputs that trigger alternative execution branches.

\section{Future Work}

In this paper, we have undertaken an extensive quantitative exploration into a large-scale MSA at Alibaba. Still, because of the substantial amount of data produced by large-scale MSAs, this investigation is not exhaustive and raises open questions that warrant future investigation. For instance, we found that the degree of MSs varies in time (Section \ref{sec:dynamicity_temporal_dependencies}), however, we did not investigate whether other local dependency features change, nor if the global topology changes, in time. Our analysis of call graph variability (Section \ref{sec:heterogeneity_call_graph_variability}) does not account for call graph structure, potentially underestimating true variability. Our findings indicated that the MSs with the most total calls, number of instances, and dependencies (Section \ref{sec:heterogeneity_relationships}) were generally not the same, however, there are other aspects of the dataset that may help elucidate links between these groups. To encourage further exploration of this dataset by the research community and to ensure replicability, we will release all code and processed data used in our analysis upon acceptance of this paper.

Overall, our analysis has highlighted similarities and differences between this Alibaba deployment and Meta's, including surprising similarities in global and local dependency structures, MS inequality, and architectural evolution. However, this work only compares between two MSAs due to the limited access to datasets from commercial entities. Therefore, we call for both future empirical explorations of other real-world MSAs, and for greater open accessibility to data, to aid in building our understanding of common principals and idiosyncrasies of these complex systems.

\bibliographystyle{ACM-Reference-Format}
\bibliography{refs}

\end{document}